\documentclass[prb,superscriptaddress,twocolumn,amsmath,amssymb,floatfix]{revtex4-1}

\usepackage{latexsym}
\usepackage{dcolumn}
\usepackage[dvips]{graphicx}
\usepackage{amssymb}
\usepackage{graphicx}
\usepackage{psfrag, color}
\usepackage{verbatim}
\usepackage{amsmath}
\usepackage{epsf}

\newcommand{\ve}{{\varepsilon}}

\begin{document}
\draft

\title{Density dependent electrical conductivity in suspended
  graphene: Approaching the Dirac point in transport} 
\author{S. Das Sarma}
\affiliation{Condensed Matter Theory Center, Department of Physics, 
	 University of Maryland,
	 College Park, Maryland  20742-4111}
\author{E. H. Hwang}
\affiliation{Condensed Matter Theory Center, Department of Physics, 
	 University of Maryland,
	 College Park, Maryland  20742-4111}
\affiliation{SKKU Advanced Institute of Nanotechnology, Sungkyunkwan
  University, Suwon 440-746, Korea } 
\date{\today}

\begin{abstract}
We theoretically consider, comparing with the existing
experimental literature, the electrical conductivity of gated monolayer graphene
as a function of carrier density, temperature, and disorder in order
to assess the prospects of accessing the Dirac point using transport
studies in high-quality suspended graphene. We show that the
temperature dependence of graphene conductivity around the charge
neutrality point provides information about how close the system can
approach the Dirac point although competition between long-range and
short-range disorder as well as between diffusive and ballistic
transport may considerably complicate the picture. 
We also find that acoustic phonon scattering contribution to the
graphene resistivity is always relevant at the Dirac point in contrast
to higher density situations where the acoustic phonon contribution to
the resistivity is strongly suppressed at the low temperature
Bloch-Gr\"{u}neisen regime. 
We provide detailed
numerical results for temperature and density dependent conductivity
for suspended graphene.
\end{abstract}

\maketitle

\section{introduction}

It has been
shown\cite{du2008,du2009,bolotin2008,bolotin2009,mayorov2012,elias2011} that
suspended graphene can achieve very high 
mobilities since various annealing techniques can remove much of the
extrinsic impurities unavoidably present in graphene on
substrates\cite{novoselov2004,tan2007,chen2008}. Such ultrapure
graphene (in this article, 'graphene' would 
mostly imply 'suspended graphene' without any substrates) with
ultrahigh mobility is of considerable importance for a number of
reasons. First, a careful comparison between graphene experimental
data with and without substrates could inform the community a great
deal about the type of disorder operational in graphene on various
substrates and the associated scattering mechanisms limiting graphene
mobility on substrates \cite{dassarma2011}, thus helping the eventual technological
application of graphene based devices. Second, ultrapure suspended
graphene enables the study of interaction effects
\cite{elias2011,dassarma2007,gonzalez1994,khvesh2001} without the
considerable complications arising from disorder, and additionally, the
absence of dielectric screening by the substrate enhances the Coulomb
interaction. Third, as a direct consequence of the high sample purity,
suspended graphene is a convenient system for the experimental study
of the fractional quantum Hall effect
\cite{du2009,bolotin2009}. Fourth, the (relative) absence 
of disorder in suspended graphene (SG) suppresses the electron-hole
puddle formation around the charge-neutrality point (CNP), i.e., the
Dirac point, making it relatively easier to access the intrinsic Dirac
point physics by lowering the carrier density \cite{mayorov2012}. 

The last item
(``accessing the Dirac point'') along with the first item
(``understanding transport in SG'') provide the motivation of our
theoretical research presented in the current paper, where we carry out
a detailed quantitative study of SG carrier transport as a systematic
function of temperature and carrier density neglecting all
complications arising from the inhomogeneous electron-hole puddles
\cite{martin2008,zhang2009,xue2011} in
the system (which typically become operational below a typical carrier
density $n_c \sim 10^{12}$ cm$^{-2}$ for graphene on SiO$_2$
substrates\cite{martin2008,zhang2009}). Our theoretical results
provide a direct estimate of the 
disorder-limited SG transport properties down to low carrier densities
which experiments should be able to access in clean SG samples
where inhomogeneous puddle formation is pushed down to very low
carrier densities. Our
work, therefore, should provide a benchmark for understanding SG
transport data as well as for figuring out how close to the Dirac
point specific SG experimental samples manage to approach.

A characteristic and universal feature of graphene transport is the
minimum conductivity phenomenon where at some disorder-dependent low
carrier density ($n_c$) the conductivity shows an approximate
saturation as a function of carrier density, forming a rough minimum
conductivity plateau around the Dirac point
\cite{novoselov2004,tan2007,chen2008} with a characteristic
electron-hole density width of $\pm n_c$. The characteristic density
cut-off $n_c$ defining this minimum conductivity plateau roughly
defines how close in density [or in energy $\varepsilon_c \approx
\varepsilon_F(n_c)=\hbar 
v_F \sqrt{\pi n_c}$, where $\varepsilon_c$ is the graphene Fermi energy for carrier
density $n_c$, and $v_F$ is the graphene Fermi velocity] the
particular graphene sample approaches the Dirac point. Larger the
$n_c$ (or $E_c$), further the system is from the Dirac point no matter
how low one tunes the gate voltage since the Dirac point is defined
only to the uncertainty of $n_c$. It is now reasonably
well-established\cite{tan2007,chen2008,dassarma2011,hwang2007,adam2007} that
the minimum conductivity plateau and the 
characteristic density cut-off arises from disorder-induced density
inhomogeneity (or equivalently electron-hole 
puddle formation) in the system which makes it impossible to access
the Dirac point nominally existing precisely at zero carrier
density. Instead the disorder-induced density fluctuations
characterized by $n_c$ make the zero-density (and as such,
measure-zero) Dirac point ill-defined over a scale of $n_c$. Smaller
$n_c$ is, closer one can approach the Dirac point by tuning the gate
voltage induced carrier density. Thus, transport measurements, which
probe $n_c$ directly by definition (since for $|n| <n_c$ the
conductivity approximately saturates) provide a clear signature for
how close to the Dirac point one is able to approach in a particular
graphene sample. In high-quality SG, $n_c \sim 10^8$ cm$^{-2}$ can be
achieved\cite{mayorov2012}, indicating that experiments can assess the Dirac point
within $\varepsilon_c \sim 0.4meV \sim 5K$. With further improvement in SG
sample quality, it is conceivable that the SG Dirac point could be
accessed within 0.5K leading to the possibility of studying intrinsic
interaction phenomenon associated with the non-Fermi liquid aspects of
the Dirac point \cite{dassarma2007,gonzalez1994,khvesh2001}. This
Dirac point accessibility is the primary 
motivation for our detailed current study of SG transport properties
as functions of carrier density and  temperature.
In this work we assume $n_c=0$, and our results therefore only apply to
ultrapure SG samples at doping densities above the conductivity minima
and/or at $k_BT>\varepsilon_c$. 

In addition to the Dirac point accessibility issue discussed above, a
secondary motivation of our work is a qualitative theoretical
understanding of realistic SG transport in order to assess whether the
current experimental SG samples are in the ballistic or the diffusive
regime. Several recent SG experimental investigations
\cite{du2008,du2009,bolotin2008,bolotin2009,mayorov2012} conclude that
their studied samples are in the ballistic regime based on the
estimated transport mean free path being longer than (or comparable
to) the linear sample size. Such very long mean free paths imply
essentially no carrier scattering within the sample (and consequently,
almost no disorder), and thus the issue of diffusive versus ballistic transport
in SG samples is an important topic of considerable interest to the
community. We find that this is also a very subtle topic since the
extraction of the mean free path from the measured conductivity is
quite nontrivial at low gate voltage (i.e., near the Dirac point) where
intrinsic thermal carrier occupancy (because of the zero band gap
nature of graphene) effects become crucial, and a n\"{a}ive estimate
of the mean free path using simply the gate-induced carrier density
would seriously over-estimate the mean free path. In fact, we believe
that at low carrier densities it is much more sensible to discuss the
physics simply in terms of the dimensionless 2D conductivity (in units
of $e^2/h$) rather than in terms of the mean free path and/or mobility
which are both derived by dividing the measured conductivity by a
putative carrier density subject to large errors near the CNP. We find
that a theoretical description based on purely diffusive transport
using the semi-classical Drude-Boltzmann theory gives a reasonable
description for the experimentally observed SG transport
properties. We believe that the only way to definitively establish
ballistic SG transport is to experimentally observe the explicit
sample size dependent conductivity
characterizing ballistic transport where conductance and not
conductivity is the meaningful physical quantity, which, to the best of our
knowledge, has not yet been seen in any SG samples by any experimental
group. We therefore contend, based on our theoretical results, that
the currently existing SG samples are all high-mobility diffusive
samples.

We consider primarily disorder-induced resistive scattering in our
theory \cite{dassarma2011,hwang2007,adam2007} since our interest is
mainly the issue of approaching the Dirac 
point in high-quality SG. The phonon effects have been considered
elsewhere in details
\cite{hwang2008,kaasbjerg2012,chen2008b,efetov2010}, and it is 
straightforward to include phonons in 
the theory, 
and we do provide some results including phonon scattering in the
theory since their effect could be important at higher (lower)
temperatures (carrier densities).

The rest of this article is organized as follows. In section II we
describe our basic transport model and provide the expected
theoretical results for finite temperature Drude transport of intrinsic
(i.e. undoped) graphene precisely at the Dirac point, which serves as
the starting point for later discussions. In section III we provide our
full theory, and then in section IV we provide our numerical results,
concluding in section V with a discussion and a summary.

\section{intrinsic transport at the Dirac point}

Precisely at the Dirac point ($n=0$), assuming no disorder-induced
electron-hole puddles and $T=0$, it is easy to see that the
semiclassical Drude-Boltzmann conductivity $\sigma_D$ at the Dirac
point (or equivalently, the charge neutrality point) is precisely zero
(i.e. infinite resistivity) because of the trivial reason that there
are no carriers to carry any current. 
We note that in our zeroth-order Drude-Boltzmann transport theory the
matrix element of 
the off-diagonal terms vanishes due to the conservation of energy,
which gives rise to the zero conductivity at T=0 and n=0. But more
rigorous transport theories (such as Kubo formula or self-consistent
Boltzmann transport theory which are beyond the Boltzmann theory) produce the
non-vanishing matrix elements between off-diagonal terms even at
n=0. Thus the well known minimum conductivity appears in these
theories. Since our analysis is totally based on the zeroth order
Boltzmann theory the conductivity vanishes for n=0 even for chiral
graphene.  
This trivial result is unstable
because there will be a finite conductivity the moment the carrier
density deviates from the precise measure-zero $n=0$ constraint which
is bound to happen at $T \neq 0$ even at the Dirac point by virtue of the
low-energy thermal electron-hole excitations capable of carrying the
current. Unlike ordinary band insulators with finite band gaps
\cite{ando1982}, there 
is no exponential suppression of finite-temperature band conductivity
in graphene because of its gaplessness. Instead, as is well-known and
discussed in some details below, $\sigma_D(T)$ at the Dirac point of
graphene manifests a power law ``insulating'' temperature dependence,
which should distinguish the Dirac point behavior from the saturated
conductivity behavior in the presence of electron-hole puddles
\cite{adam2007,li2011}. 
This power-law ``insulating'' behavior associated with the Dirac point
has nothing to do with Anderson localization physics and arises
entirely within the metallic Drude-Boltzmann diffusive transport
theory because graphene is a  gapless semiconductor.

Consider undoped graphene in the absence of disorder (or electron
hole puddles), i.e., the chemical potential at $T=0$ lies at the Dirac point.
Then the thermally excited number of electrons (and holes) at finite
temperatures can 
be calculated  
\begin{equation}
n= \int D(\varepsilon) n_F(\varepsilon) d \varepsilon,
\label{den_t0}
\end{equation}
where $n_F$ is the Fermi distribution function and $D(\varepsilon) = g
\varepsilon/2\pi \gamma^2$ is the density of 
states of graphene with the total degeneracy $g=4$ arising from spin
(2) and valley (2)and $\gamma = \hbar
v_F$. The induced carrier density at $T$ becomes
\begin{equation}
n  = \frac{g}{2\pi}\frac{\pi^2}{12} \frac{(k_BT)^2}{\gamma^2} 
 =  T^2 \times 0.89 \times 10^6  \; {\rm cm}^{-2}.
\end{equation}
At $T=300K$ we have $n=8\times10^{10}$ cm$^{-2}$. Thus if the
conductivity is simply proportional to the carrier density, it
increases quadratically with temperature. 
We show below that the actual temperature dependence of conductivity
depends on the scattering mechanism.

In the presence of disorder induced momentum scattering the
conductivity can be calculated within 
Boltzmann transport theory. In this theory the puddle effect is not
considered i.e., the theory is valid only for $n >n_c$.  The
conductivity is given by  \cite{dassarma2011,hwang2009}
\begin{equation}
\sigma_D(T) = \frac{e^2 v_F^2}{2} \int d \varepsilon D(\varepsilon)
\tau(\varepsilon) \left ( -\frac{\partial f (\varepsilon)}{\partial
  \varepsilon} \right ),
\label{eq:sigma_B}
\end{equation}
where $\tau$ is the disorder induced transport scattering time.
Note that in this equation the conductivity is not related explicitly to
the carrier density. If we assume a constant scattering time, i.e. no
energy and temperature dependence of the scattering time, then we
have,
\begin{equation}
\sigma_D(T) = \frac{e^2}{h}\frac{2 \ln 2}{ \gamma} (\tau_0v_F) (k_BT),
\end{equation}
where $\tau_0$ is the constant scattering time  and the mean free path
is given by $l = \tau_0 v_F$. In this case the conductivity
increases linearly with temperature. 

Now consider a generalized scattering time. Within the Fermi golden rule
we have
\begin{equation}
\frac{1}{\tau} = \frac{2\pi}{\hbar} n_i \sum_{k'} \left | V_i(k,k') \right
|^2 (1-\cos \theta_{kk'}) \delta(\varepsilon_k - \varepsilon_{k'}),
\end{equation}
where $n_i$ is the impurity concentration and $V_i$ is the
carrier-impurity scattering potential. 

For a short range potential (i.e. $\delta$-range potential) with the
strength $V_i=V_{\delta}$ and the impurity density $n_i = n_{\delta}$ we have
\begin{equation}
\frac{1}{\tau(\varepsilon)} = \frac{n_{\delta}V_{\delta}^2}{4} \frac{v_F
  \varepsilon}{\gamma^3}. 
\end{equation}
With Eq.~(\ref{eq:sigma_B}) we have
\begin{equation}
\sigma_D(T) = \frac{4 e^2}{h} \frac{\gamma^2}{n_{\delta}V_{\delta}^2}.
\end{equation}
Thus the conductivity is independent of the temperature for
$\delta$-correlated zero-range disorder.

For unscreened long ranged Coulomb potential, $V_i(q) = 2\pi
e^2/\kappa q$ where $\kappa$ is the background lattice dielectric
constant (taken to be unity for SG), we have
\begin{equation}
\frac{1}{\tau(\varepsilon)} = \frac{\pi^2}{\hbar} n_i \frac{r_s^2
  \gamma^2}{\varepsilon},
\end{equation}
where $r_s = e^2/\kappa \gamma$ is the graphene fine structure
coupling constant. With Eq.~(\ref{eq:sigma_B}) we have
\begin{equation}
\sigma_D(T) = \frac{e^2}{h} \frac{1}{3n_i} \frac{1}{r_s^2 \gamma^2}
(k_BT)^2
\label{eq:sigma_dt}
\end{equation}

For screened long range Coulomb potential, i.e.,
\begin{equation}
V_i(q) = \frac{2\pi e^2}{\kappa q \epsilon(q)},
\end{equation}
where the dielectric function, $\epsilon(q)$, is given by
\begin{equation}
\epsilon(q) = 1 + \frac{2\pi e^2}{\kappa q} \Pi(q,T),
\end{equation}
where $\Pi(q)$ is the polarizability depending on the wave vector and
temperature\cite{hwang2007a}. Within RPA we have,
\begin{eqnarray}
\Pi(q,T& ) &  =   \frac{q}{4 \gamma} \nonumber \\
 &+ &  \frac{4}{\pi \beta \gamma^2}  \left [ \ln 2 - \int _0^{\beta
    \varepsilon_q/2} \frac{\sqrt{1-(2y/\beta \varepsilon_q)^2}}{1+e^y}
      dy \right ],
\end{eqnarray}
where $\beta = 1/k_BT$ and $\varepsilon_q=\hbar v_F q$. Finally, the
conductivity can be calculated to 
be (with some straightforward algebra)
\begin{equation}
\sigma_D(T) = \frac{e^2}{h}\frac{1}{2\pi n_i} \frac{(k_BT)^2}{r_0^2 \gamma^2}
I(r_0),
\label{eq:sigma_ddt}
\end{equation}
where
\begin{equation}
r_0 = \frac{r_s}{1 + \pi r_s /2},
\end{equation}
and $I(r_0)$ is a function which is independent of the temperature and
given by
\begin{equation}
I(r_0) = \int_0^{\infty}dt \; t^2 \tau(t,r_0) \frac{e^t}{(e^t + 1)^2},
\end{equation}
where
\begin{equation}
\frac{1}{\tau(t,r_0)} = \int_0^1 dx \frac{\sqrt{1-x^2}}{\epsilon_0(2tx,r_0)^2} 
\end{equation}
and
\begin{equation}
\epsilon_0(z,r_0) = 1 + \frac{4 r_0}{z} \left [ \ln 2 - \int_0^{z/2}
  \frac{\sqrt{1-(2y/z)^2}}{1+e^y} dy \right ].
\end{equation}
Thus the Dirac point conductivity increases quadratically with
temperature for screened
Coulomb potential disorder similar to the bare Coulomb disorder
results in Eq.~(\ref{eq:sigma_dt}). 

Finally, for scattering of the thermally excited carriers by the
deformation potential coupling to the acoustic phonons\cite{hwang2008} we get the
following expression in the high-temperature nondegenerate
equipartition phonon distribution regime:
\begin{equation}
\sigma_{ph} = \frac{e^2}{h}\frac{8\rho_m v_{ph}^2\gamma^2}{D^2}
\frac{1}{k_B T},
\label{eq:sigma_ph}
\end{equation}
where $D$ is the deformation potential, $\rho_m$ the graphene mass
density, and $v_{ph}$ the phonon velocity. Thus, the conductivity
decreases inverse linearly with increasing temperature.
For low temperatures, $T<T_{BG}$ where the $T_{BG}$ is the so-called
Bloch-Gr\"{u}neisen temperature, phonon scattering is very strongly
suppressed\cite{hwang2008} and is not of any interest in the current
work.

We note that, as expected, the above Boltzmann theoretical
semiclassical description of the Dirac point conductivity, which
neglects all interactions \cite{muller2008} and interference effects
\cite{aleiner2006,wu2007,morozov2006} (but includes
thermal excitation, screening, and scattering effects
quantum-mechanically), gives $\sigma_D(T=0) = 0$ at the Dirac point,
and the finite $\sigma_D(T)$ for $T\neq 0$ arises entirely from the
finite density of the thermal electron-hole excitations [c.f.,
  Eq.~(\ref{den_t0})] in gapless graphene. The temperature dependence of the
finite-temperature Dirac point conductivity is entirely a power law
with $\sigma_D(T) \sim T^{\alpha}$ where $\alpha = 0$, 1, 2
respectively depending on whether the scattering mechanism is
short-ranged or energy-independent or long-ranged (including screened
Coulomb scattering). 
In addition, $\alpha=-1$ for phonon scattering as shown in Eq.~(\ref{eq:sigma_ph}), and
in the presence of all possible scattering mechanisms, the actual
temperature dependent Dirac point conductivity would be nonuniversal
and complex, depending on the strength of the various scattering
processes in the particular sample.
It is then easy to see that the experimentally
measured $\alpha$ exponent could be any number between 0 and 2,
depending on the manifestly non-universal strength of various
scattering mechanisms in the system.

\begin{figure}[t]
	\centering
	\includegraphics[width=1.0\columnwidth]{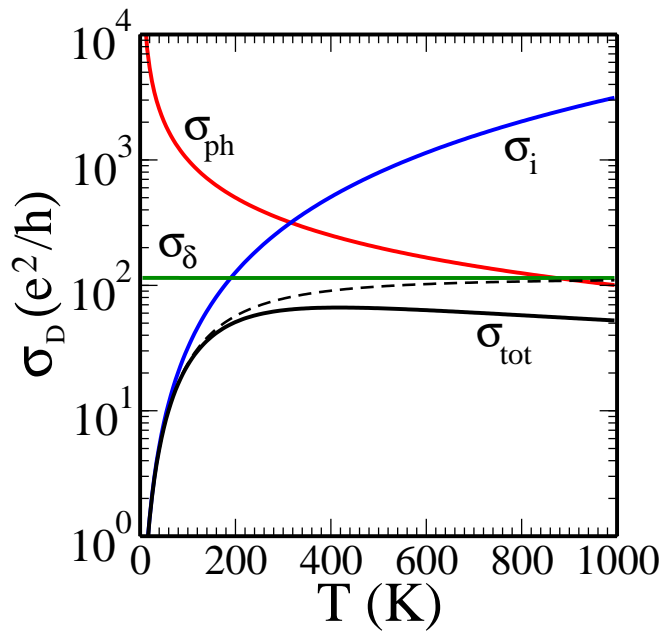}
	\caption{
(color online) The Dirac point conductivity $\sigma_D$ as a function of
temperature. $\sigma_i$ ($\sigma_{\delta}$) indicates the conductivity
due to screened charged Coulomb disorder with an impurity density $n_i=0.3\times
10^{10} cm^{-2}$ [short-range disorder with $n_{\delta}V_{\delta}^2 =
1.5 (eV\AA)^2$]. $\sigma_{ph}$ represents the conductivity limited by
acoustic phonon scattering. $\sigma_{tot}$ is the total conductivity
including all scattering mechanisms. The dashed line indicates the
conductivity due to the Coulomb disorder and the short-range disorder.  
\label{fig:fig0}
}
\end{figure}

The important point to note is that the temperature dependence is
never exponential, a key qualitative feature which helps to
distinguish the Dirac point thermally induced conductivity from the
Anderson (strong) localization (or gap-induced insulating)
behavior. This power-law temperature-dependence feature remains valid even in the presence of
phonon scattering which leads to a metallic conductivity (i.e., a
negative $\alpha$) with a temperature-dependent conductivity with a
power law between 1 and 5,\cite{hwang2008} depending on whether Bloch-Gr\"{u}neisen
regime is relevant or not. We emphasize that phonons can only induce
metallic behavior (with the conductivity decreasing with increasing
temperature), and as such disorder and phonon scattering together may
produce a complicated non-monotonic temperature dependence.
In Fig.~\ref{fig:fig0} we show the calculated $\sigma_D$ including all
three scattering mechanisms (i.e. short-range and long-range disorder
as well as acoustic phonons).
The total conductivity ($\sigma_{tot}$)
shows the non-monotonic temperature dependence. As temperature
increases, $\sigma_{tot}$ increases due to the dominance of Coulomb
disorder at low temperatures, but after reaching a maximum
conductivity it decreases with increasing
temperature due to phonon scattering. 
This crossover temperature scale for $\sigma_D (T)$ depends sensitively
on the amount of Coulomb disorder in the system, and increases
(decreases) with increasing (decreasing) Coulomb disorder.  We note
that if Coulomb disorder is weak or absent, $\sigma_D (T)$ decreases
monotonically with increasing temperature because of phonon
scattering.  The fact that experimental low-density SG transport data
\cite{du2008,du2009,bolotin2008,bolotin2009,mayorov2012} 
show a nonmonotonic temperature
dependence in the low-density SG conductivity clearly indicates that
Coulomb disorder dominates even the currently existing SG samples (and
not just the graphene on substrate samples).

A key aspect of the temperature-dependent Dirac point electrical
conductivity derived above, which, although rather obvious, has not
been much discussed in the literature, is that the intrinsic Dirac
point behavior is really a {\it high-temperature} phenomenon rather
than a low-temperature one since one must have $n(T) > n_c$ in order
to see the intrinsic behavior (where $n_c$ is the characteristic
cut-off density defining electron-hole puddle formation in the
system). Thus, the intrinsic Dirac point physics can only be accessed
for $T \gg T_c \approx 10^{-3} \sqrt{n_c}$ with $n_c$ (T) measured in
units of cm$^{-2}$ (K), and the intrinsic Dirac point behavior is
completely suppressed by the extrinsic inhomogeneous carrier density
fluctuations associated with the electron-hole puddles. For the
extremely low value of $n_c \sim 10^8$ cm$^{-2}$, we get $T_c = 10K$
whereas for the usual graphene on SiO$_2$ substrates, where $n_c
\approx 10^{12}$ cm$^{-2}$, $T_c \approx 1000K$! Thus, the intrinsic
Dirac point conductivity (and its strongly insulating temperature
dependence arising from Coulomb disorder) can never be observed in
most graphene on SiO$_2$ samples 
studied in most laboratories, and indeed, in spite of clear
theoretical predictions for the insulating temperature dependence of
the low-density graphene conductivity\cite{hwang2009}, for a long time it was believed
that graphene conductivity is essentially temperature independent upto
room temperature (since the electron-phonon coupling constant is small
in graphene, even phonon-induce metallic temperature dependence is
fairly weak in graphene at high carrier density).

To observe the intrinsic Dirac point physics $\sigma_D(T)$ at low
temperatures ($\alt 100$ mK) so that various predicted interaction
induced Dirac point reconstruction (or instability) \cite{khvesh2001} can be
experimentally observed (since higher temperature strongly suppresses
interaction effects), one would have to produce SG samples of rather
extraordinary purity with the puddle induced density inhomogeneity
being less than $10^4$ cm$^{-2}$. This seems to be a rather daunting
task, and it is therefore safe to say that the $T\rightarrow 0$
intrinsic Dirac point conductivity is unlikely to be experimentally
explored in the near future, making our current work, where we
consider finite-temperature Dirac point transport neglecting
interaction effects, to be relevant for all experimental Dirac point
transport studies in the near future.

Before concluding this section we emphasize that we are only
considering $T \neq 0$ disorder-limited Boltzmann conductivity in our
theory, neglecting all interaction effects, and for $T=0$ our Dirac
point conductivity is trivially zero. A completely different approach
is necessary to discuss intrinsic Dirac point conductivity in clean
graphene at $T=0$ where interaction and quantum interference effects
would be important. Such a theory is beyond the scope of our work and
is not of interest to us since we know of no experimental relevance of
the $T=0$ Dirac point conductivity. Second, the inclusion of phonon
effects is 
extremely important for the Dirac point conductivity behavior as a
function of temperature since phonons lead to the 
non-monotonic $\sigma_D(T)$ with metallic behavior ($d\sigma_D/dT <
0$) at higher temperatures replacing the insulating behavior
($d\sigma_D/dT >0$) at lower temperatures with a nonuniversal
disorder-dependent crossover behavior.

\section{conductivity of suspended graphene}

Our theoretical model assumes the absence of puddles in the system, and
as such, the theory is cut off at some samples-dependent
characteristic carrier density $n_c$ below which the inhomogeneous
density fluctuations around the Dirac point become important, leading
to an observable minimum conductivity plateau formation. The theory is
valid only for $n \agt n_c$ or in the non-plateau regime by
definition. If $n_c$ is very small, as has been claimed in several
recent experimental studies
\cite{du2008,du2009,bolotin2008,bolotin2009,mayorov2012,elias2011}, 
then our theory would apply down to very
low carrier density (as long as interference and interaction effects
are negligible). Puddle effects on graphene transport properties have
been theoretically studied elsewhere \cite{li2011}, and puddles would introduce
additional nontrivial temperature and density dependence for $n \alt
n_c$, which is not of interest to us in the current work where $n_c$
is very small by virtue of the ultra-clean nature of SG in general.

In our theory, we consider three distinct scattering mechanisms
contributing to the SG resistivity. These are charged impurity,
short-range disorder, and in-plane acoustic phonon scattering
processes. There
can be other types of scattering mechanisms contributing to the
graphene resistivity such as resonant scattering centers
\cite{stauber2007,ferreira2011}, ripples \cite{fasolino2007}, and
flexural phonons \cite{mariani2008,gornyi2012,mariani2010}. 
The short-ranged disorders considered in the resonant scatterers
\cite{stauber2007,ferreira2011} 
modify the density of states of graphene (i.e., there are a resonant
DOS peak at the Dirac point due to the disorders). In this case the
short-ranged disorder scattering gives rise to the
density dependent conductivity. In our Boltzmann
theory, to keep the theory to be
consistent for all other disorders we used the bare DOS
of pure graphene and we have the density independent
conductivity from the short-range disorder.
In this paper the short-range disorder represents the zero-range
disorder, i.e., $V(r)=V_{\delta} \delta(r)$. 
The calculated scattering time with the finite
width potential (i.e., $V(r)=V_{\delta} \theta(r-r_0)$)
does not significantly modify that with the zero-range potential
as long as $r_0 < 2a$ where $a$ is the lattice constant of graphene.   
In suspended graphene the flexural phonons may dominate the phonon
contribution to the resistivity
\cite{gornyi2012,mariani2010}. However, in suspended graphene under 
specific tension induced by 
contacts (this is the case for all available suspended graphene
samples) the flexural phonon contribution to the conductivity is
severely suppressed and as a consequence the in-plane phonon is the
dominant scattering mechanism \cite{mariani2010}. Since we consider
suspended graphene under the tension our calculated results is not affected by
the flexural phonons. 
The possibility of still other unknown scattering mechanisms (such as
scattering from the hybridization of electron-hole excitations and
out-of-plane optical phonons \cite{badalyan2012}) contributing to the
graphene resistivity 
cannot be ruled out either. 
But we neglect these scattering mechanisms because we
want to keep the number of parameters to a minimum and also because
the very high SG mobility and quality imply that the overall
scattering contributions are small.

The Drude-Boltzmann conductivity theory for extrinsic graphene in the
presence of induced carriers is a straightforward generalization of the
theory provided in Sec. II except the total carrier density now has an
externally tunable (through the gate voltage) density in addition to
the thermally excited intrinsic carriers considered in Sec. II. This
theory has been much discussed in the literature, and we provide below
the working equations for different contributions to the SG
resistivity from the three different scattering mechanisms considered
in our work. The theory below is a straightforward generalization of
the theory for the finite-temperature Dirac point conductivity
developed in the last section.

Within the Boltzmann transport theory \cite{dassarma2011,hwang2009},
the conductivity $\sigma(n,T)$ 
is given within the relaxation time approximation by
\begin{equation}
\sigma =\frac{e^2}{2}\int d\epsilon D(\epsilon) v_k^2 \tau(\epsilon)
\left [ - \frac{\partial f(\epsilon)}{\partial \epsilon} \right ],
\end{equation}
where $f(\epsilon)$ is the relevant distribution function. The
relaxation time $\tau(\epsilon) \equiv \tau (\epsilon_k)$ is given after
ensemble averaging over random disorder configuration by 
\begin{eqnarray}
\frac{1}{\tau^{(\alpha)}(\epsilon_k)} = \frac{2\pi}{\hbar}
\sum_{(\alpha)} & n_i^{(\alpha)} & \int \frac{d^2k'}{(2\pi)^2} \left |
\langle V_{kk'}^{(\alpha)} \rangle \right |^2 \nonumber \\
& \times & (1-\cos \theta_{kk'}) \delta(\varepsilon_k - \varepsilon_k'),
\end{eqnarray}
where $\theta_{kk'}$ is the scattering angle and $V^{(\alpha)}$ is the potential
disorder causing the scattering with $n_i^{(\alpha)}$ being the 2D
density of the random impurities (or defects) producing the disorder
and '$(\alpha)$' being a label indicating the kind of scatterer
(e.g. long-range Coulomb scattering, short-range defect scattering,
etc.) under consideration with each scattering mechanism being
independent.

The finite-temperature conductivity is given by an appropriate thermal
energy averaging within the Boltzmann theory once $\tau(\epsilon)$ has
been calculated. The zero-temperature result is simply given by:
\begin{equation}
\sigma = \frac{e^2 v_F^2}{2} D(\epsilon_F) \tau(\epsilon_F),
\end{equation}
where the graphene Fermi velocity $v_F$ is assumed to be a constant
(independent of momentum and density) and $\epsilon_F$, the Fermi
energy, is the chemical potential at $T=0$. The finite temperature
chemical potential, $\mu(n,T)$, is calculated self-consistently\cite{hwang2009} so
that the net carrier density (induced by doping or an external gate)
is $n$, and the gaplessness of graphene automatically ensures that
this procedure incorporates the thermally excited carriers (i.e., the
only carriers present for intrinsic graphene as considered in Sec. II
at the Dirac point with $\epsilon_F=0$) along with the induced
carriers of density $n$. In this paper our main interest is the low
carrier density regime where $n$ is small so that the Dirac point
behavior is accessed.

\subsection{Short-range disorder}

For short-range (or more appropriately, zero-range) delta scatterers,
we have
\begin{equation}
|\langle V_{kk'} \rangle |^2 = V_{\delta}^2 (1+\cos \theta)/2,
\end{equation}
where $V_{\delta}$ is the strength of the short-range disorder and the
$(1+\cos \theta)/2$ factor arises from the matrix elements effect due
to the pseudospin chirality of graphene (this chirality factor leads
to the famous suppression of back scattering in graphene and also in
surface states of topological insulators).

It is easy to show that short-range disorder leads to a carrier
density independent conductivity $\sigma(n) \propto V_{\delta}^{-2}$, and to
an exponentially suppressed temperature dependence at low
temperatures. In the high temperature limit, the resistivity due to
short-range disorder increases by a factor of 2 compared with the
$T=0$ value \cite{hwang2009} and thus, short-range disorder by itself introduces weak
metallic behavior in graphene with little temperature dependence at
low ($T \ll T_F = \epsilon_F/k_B$) temperatures and
increasing resistivity at high temperatures ($T \rightarrow \infty$).
This is the same as what happens to just the Dirac point conductivity
as discussed in section II. 

\subsection{Long-range Coulomb disorder}

Unintentional charged impurity centers in the environment are a major
source of disorder for graphene on substrates. Although they are
substantially removed in annealed SG samples (leading to the very high
observed SG mobility), there are still some remnant random charged
impurity centers on the SG surface which contribute to carrier
scattering. For Coulomb disorder we have
\begin{equation}
|\langle V_{kk'} \rangle |^2 = \left | \frac{V_c(q)}{\epsilon(q)}
\right |^2 \frac{1+\cos \theta}{2},
\end{equation}
where $V_c(q) = 2\pi e^2/\kappa q$, with $\kappa$ ($=1$ for suspended
graphene) as the background dielectric constant, is the 2D Coulomb
interaction and $\epsilon(q)$ is the wave vector dependent static
dielectric function of the free carriers in graphene \cite{dassarma2011,hwang2007}.

The density dependence $\sigma(n)$ of conductivity due to Coulomb
disorder is linear, $\sigma \sim n$, and the preponderance of the
observed linearity of $\sigma(n)$ on $n$ is considered to be strong
evidence for the importance of charged impurity scattering in
determining graphene transport properties. The temperature dependence
due to Coulomb disorder has been discussed elsewhere \cite{hwang2009}, and here we
summarize the main findings for the discussion of our results
presented in the rest of this paper. In the low temperature limit ($T
\ll T_F = \epsilon_F/k_B$), one gets for Coulomb disorder
\begin{equation}
\sigma(T) = \sigma_0 \left [ 1 - A_2 (T/T_F)^2 \right ],
\label{eq:sigma_low}
\end{equation}
where $A_2 >0$. In the high temperature limit ($T \gg T_F$), which is
the more appropriate regime for our consideration of transport near
the Dirac point ($n \sim 0$), one gets
\begin{equation}
\sigma(T) \sim B_2 (T/T_F)^2, \;\;\; {\rm with} \; B_2 >0.
\label{eq:sigma_high}
\end{equation}
Thus, Coulomb disorder by itself predicts weak metallic behavior for
$T \ll T_F$ and strong insulating behavior for $T \gg T_F$ (which is
the appropriate limit for the low density Dirac point regime).

\subsection{Acoustic phonon scattering}

In addition to the short-range and long-range disorder, which affect
the SG conductivity at all temperatures (but with distinct density and
temperature dependence in different regimes), we also include
resistive scattering by graphene acoustic phonons through the
deformation potential coupling, which is operational primarily at
higher temperatures
(except at the Dirac point where it is operational at all temperatures).
We note that the deformation potential coupling
is rather weak in graphene, and therefore, the primary (essentially,
the only) effect of phonon scattering is to introduce a weak metallic
temperature dependence with the phonon-induced resistivity $\rho_{ph} \sim
T$ at higher temperatures $T > T_{BG} \sim 2\hbar v_{ph} k_F$, where
$v_{ph}$ is the phonon velocity (i.e. speed of sound). Since $k_F
\propto \sqrt{n}$, phonon effects could affect the net SG resistivity
at fairly low temperatures for low carrier density systems of our
interest in this work.
Since $k_F$ effectively vanishes at the Dirac point, acoustic phonons
are operational even at arbitrarily low temperatures near the Dirac
point as $T_{BG}$ tends toward zero.

Since phonon scattering has already been considered in details
theoretically elsewhere \cite{hwang2008} we show below the relevant
``high-temperature'' relaxation time for the deformation potential
coupling:
\begin{equation}
\frac{1}{\tau(\epsilon)} =
\frac{1}{\hbar^3}\frac{\epsilon}{4v_F^2}\frac{D^2}{\rho_m v_{ph}^2}
(k_BT),
\end{equation}
where $D$, $\rho_m$ are respectively the deformation potential
coupling and graphene mass density. 
At very low temperatures, the phonon-induced relaxation time enters
the Bloch-Gr\"{u}neisen regime where $\rho_{ph} \sim T^4$ and is
negligibly small \cite{hwang2008}. In our numerical results presented later in this
paper, we use the full numerical solution of the Boltzmann theory for
calculating the phonon-induced resistivity which always becomes
important above (a density dependent) characteristic temperature.

\subsection{Asymptotic behavior of SG conductivity}

We now combine the contribution to $\sigma(T,n) = [\rho(T,n)]^{-1}$
from the three distinct scattering mechanisms described in subsections
A, B, C above to discuss the asymptotic density and temperature
dependence of the SG conductivity near the Dirac point.

First we establish the counter-intuitive result that the conductivity
around the Dirac point is always affected by phonon scattering
even at arbitrarily low temperatures.
Writing the effective carrier density $n(T)$ around the Dirac point as 
\begin{equation}
n(T) \approx n_0 + AT^2,
\end{equation}
where $n_0 \propto V_g$ is the gate induced extrinsic carrier density
and $AT^2$ ($\gg n_0$) is the intrinsic Dirac point thermally excited
carrier density (see Sec. II), we can define an effective Fermi wave
vector:
\begin{equation}
k_F = \sqrt{\pi(n_0 + AT^2)} \approx \sqrt{\pi A} T,
\end{equation}
where  $A=(\pi/6)(k_B/\gamma)^2$ are known $T$-independent constants. Then, the
Bloch-Gr\"{u}neisen temperature $T_{BG}$ above which phonon scattering
effects are important is given by
\begin{equation}
T_{BG} = \omega_{ph}(2k_F)/k_B = 2 (v_{ph}/v_F) \sqrt{\pi^2/6}T = dT,
\end{equation}
where $d=2v_{ph}/v_F \sqrt{\pi^2/6}$.
(We note that $T_{BG}$ is defined by the phonons with effective wave vector
of $2k_F$ since $2k_F$ typically is the most resistive scattering process
across the Fermi surface.) 
Now the condition for acoustic phonons to contribute appreciably to
the resistivity (e.g., $\rho \propto T$) is that $T > T_{BG}$, implying
$d >1$. If $d>1$, then the Dirac point resistivity remains unaffected
by phonons to arbitrarily high temperatures,
whereas $d<1$ implies that phonons contribute a linear resistivity down
to low temperatures. 
It is easy to check that
for actual graphene parameters, we find that $d \approx 10^{-3}$ and
thus $d \ll 1$ is satisfied, implying that $\sigma_D(T)$ 
is affected, in principle, by phonon scattering at all temperatures.
We can estimate the crossover temperature scale $T_c$ for the
low-density SG system to go from being 'insulating-like' dominated by
Coulomb disorder to being 'metallic-like' dominated by phonon
scattering to be $T_c \sim 2/(A_c B_p)^{1/2}$ where $A_c$ is the coefficient
for the $T^2$ dependence due to Coulomb disorder in Eq.~(\ref{eq:sigma_ddt}) and $B_p$ is
the coefficient of the linear T-inverse term due to phonon
scattering in Eq.~(\ref{eq:sigma_ph}).  It is easy to show that $T_c \sim  n_i^{1/3}$, and
thus the crossover temperature increases with increasing Coulomb
disorder in the system.  For very pure SG samples, $T_c$ could be very
low, and it is in principle possible for the low-temperature Dirac
point conductivity to show a transition from being insulating-like to
being metallic-like as a function of decreasing disorder (i.e. $n_i$),
but this is by no means a localization transition -- it is simply a
crossover behavior driven by the competition between charged
impurities and phonons.  In general, the Dirac point conductivity
would show complex non-monotonic temperature-dependent conductivity as
is obvious from this analysis and from Fig.~\ref{fig:fig0}.

For finite doping when $n_0 \gg AT^2$, the above argument does not
hold, and phonon effects on conductivity are pushed to much higher
temperature while at the same time the temperature scale for the
insulating behavior is substantially suppressed since $T_F$ is now large
[see Eqs.~(\ref{eq:sigma_low}) and (\ref{eq:sigma_high}) above].
Then, the high-temperature behavior of $\sigma(T)$ must
always reflect the weak metallic ($d\sigma/dT <0$) conductivity in the
$\rho_{ph} \propto T$ regime for $T \gg T_{BG}$. 
Thus, at high carrier density both the insulating and the metallic
temperature dependences are strongly suppressed leading to very weak
temperature dependence of graphene conductivity as is well-established
experimentally. How high in
temperature must one go to manifest the weak metallic phonon-induced
conductivity obviously depends on the gate induced carrier
density. 
First, $T_{BG}$ increases with increasing carrier
density $n$ since $T_{BG} \propto \sqrt{n}$. 
Thus, doped SG with high carrier density should reflect very weak
temperature dependence except for the weak phonon-induced metallic
behavior at high temperatures whereas the Dirac point conductivity (or
more generally, low-temperature conductivity) should reflect strong
(and in principle, nonmonotonic) temperature dependence in the
conductivity.  An observation of any strong temperature dependence in
graphene (insulating, metallic, or nonmonotonic) therefore indicates
the Dirac point behavior.

For short-range disorder, as discussed in IIIA above, $\sigma(T)$ has
a weak T-dependence at low temperatures and metallic temperature
dependence at high temperatures, and thus adding phonon
scattering does not change the picture qualitatively. Thus, pure
short-range disorder by itself can only introduce weak metallic
temperature dependence in graphene transport properties with
$\sigma(T)$ decreasing with increasing $T$ at high temperatures as
phonons start playing a role. In addition, short-range disorder does
not manifest any density dependence in $\sigma(n)$, and therefore,
$\sigma(T,n)$ would have little dependence on density and temperature
(except at high temperatures) if the dominant resistive scattering
mechanism is short-range defect scattering in conflict with all
existing experiments.

Long-range disorder, however, must dominate low-density transport
since $\rho \propto 1/n$ for long-range disorder, and therefore, the Dirac
point conductivity is necessarily limited by long-range disorder,
which, as discussed in Sec. IIIB above, leads to a non-monotonic
temperature dependence of weak metallic behavior for
higher values of temperature. At low carrier densities (as well as in
the presence of any remnant puddles) the low-temperature weak
metallicity may be ignored, and the net temperature dependence arising
from long-range disorder (IIIB) and phonon scattering (IIIC) can be
combined to give
\begin{equation}
\rho(T) = (\sigma_0 + A_cT^2)^{-1} + B_p T,
\label{eq:rho_t}
\end{equation}
where $\sigma_0$ subsumes the weak metallic temperature dependence at
low temperatures and $A_c$, $B_p$ (as defined above) depend respectively on the
long-range Coulomb scattering and acoustic phonon scattering as
discussed above. For small $\sigma_0$, at the Dirac point,
such a
T-dependence leads immediately to a universal crossover temperature
scale $T_c$ given by
\begin{equation}
T_c \approx (AB)^{-1/3} \propto
  n_i^{1/3}r_s^{2/3}v_F^{4/3}v_{ph}^{1/3}D^{-2/3}, 
\end{equation}
with $T_c$ being the characteristic temperature defining the crossover
from the temperature power-law insulating temperature dependence
induced by Coulomb scattering to the higher-temperature phonon-induced
weak metallic temperature dependence. We note that $T_c$ increases
weakly with disorder ($n_i$) and $r_s$, but decreases with increasing
deformation potential coupling. Cleaner SG systems would thus manifest
stronger phonon effects. 
But for high densities, when $\sigma_0$ typically is large, and the
phonon effects can only show up at very high temperatures
$T \gg T_{BG}(\sim n_0)$ with $T_{BG}$ also being large,
Eq.~(\ref{eq:rho_t}) immediately implies 
very little temperature dependence, except for $\rho(T) \sim T$ for $
T \gg T_{BG}$.
Thus, away from the Dirac point, whence both $T_F$ and $T_{BG}$
are large, SG conductivity should manifest weak temperature dependence
whereas the observation of a strong temperature dependent conductivity
is evidence for approaching the Dirac point in the system. 

In the next section, we provide our calculated numerical results for
SG transport properties using the full numerical solutions of the
Boltzmann transport theory including long-range and short-range
disorder and acoustic phonon scattering.

\section{numerical transport results for suspended graphene}

We consider three different scattering mechanisms in calculating the
density and temperature dependent SG conductivity $\sigma(n,T)$, or
equivalently the resistivity $\rho \equiv 1/\sigma$: long range
Coulomb disorder ($n_i$), short range disorder
($n_{\delta}V_{\delta}^2$, where $n_{\delta}$ is the short-range impurity density
and  $V_{\delta}$ is the strength of
the short-range disorder) and acoustic phonon scattering ($D$). We
assume $D=19$ eV throughout and assume that the long range and the
short range disorder can both be taken to arise from random quenched
point impurity centers located on the suspended graphene layer.

In addition to conductivity (or resistivity) we also present results
for the mobility $\mu$ and the mean free path $l$ since these are
quantities of considerable experimental interest. In particular, the
mean free path is often used by the experimentalists to
operationally determine whether transport is ballistic or not -- if $l
> L$ (where $L$ is the system size) one nominally has ballistic
transport (and our theory becomes inapplicable). Similarly, mobility
is an important physical quantity pertaining to the sample quality --
typically SG samples should have high mobility because the amount of
disorder is suppressed.

In calculating the conductivity of extrinsic suspended graphene in the
presence of finite doping (or gate induced carriers with a finite
Fermi energy $\epsilon_F$) we first generalize the theory of Sec. II
to the finite doping case as discussed below (with $n_0$ being the
doping density).

\begin{figure}[t]
	\centering
	\includegraphics[width=1.0\columnwidth]{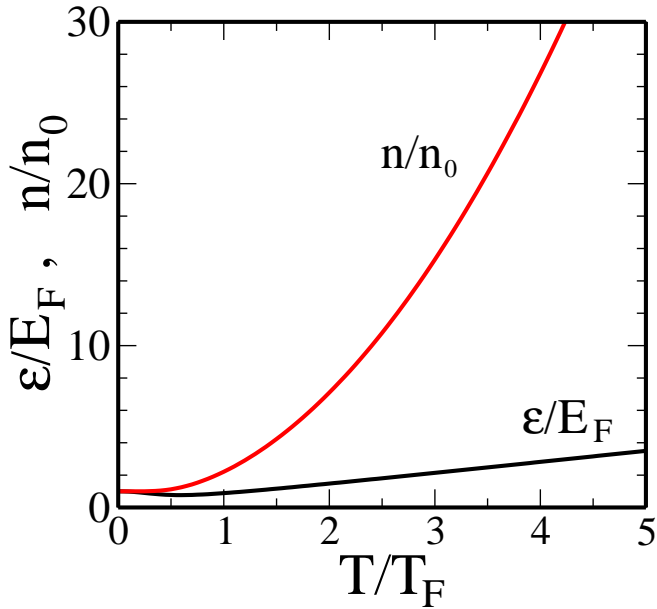}
	\caption{
(color online) Temperature dependent electron density $n(T)$
          [Eq.~(\ref{eq:nt})] and energy 
$\ve(T)$ [Eq.~(\ref{eq:et})] as a
function of temperature, $T/T_F$.
\label{fig:fig1}
}
\end{figure}

The current density in the presence of an applied electric field
($E_x$) is given by
\begin{equation}
J_x = E_x \frac{e^2 v_F^2}{2} \int D(\varepsilon) \tau(\varepsilon) \left
( -\frac{df(\ve)}{d\ve} \right ) d
\varepsilon,
\label{eq:j0}
\end{equation}
where $D(\ve) = g \ve/[2\pi (\hbar v_F)^2]$ is the density of states
of graphene with energy $\ve = \hbar v_F k$ and $f(\ve)$ is the Fermi
distribution function. Thus the conductivity becomes
\begin{equation}
\sigma =  \frac{e^2 v_F^2}{2} \int D(\varepsilon) \tau(\varepsilon) \left
( -\frac{df(\ve)}{d\ve} \right ) d \varepsilon.
\label{eq:sigma}
\end{equation}
To find a direct analogy of the conductivity with the parabolic
dispersion, $\sigma = ne^2\langle 
\tau \rangle /m$, we rewrite Eq.~(\ref{eq:sigma}) as
\begin{equation}
\sigma(T) = e^2 \langle \tau \rangle \frac{g \ve(T)}{4 \pi \hbar^2},
\label{eq:sigma0}
\end{equation}
where 
\begin{equation}
\langle \tau \rangle = \frac{\int D(\varepsilon) \tau(\varepsilon) \left
( -\frac{df(\ve)}{d\ve} \right ) d\varepsilon}{\int D(\varepsilon) \left
( -\frac{df(\ve)}{d\ve} \right ) d\varepsilon},
\end{equation}
which is exactly the same definition of the average scattering time for
2D parabolic band systems
and $\ve(T)$ is given by
\begin{equation}
\ve(T) = \int f(\ve) d \ve = \mu_0(T) + \frac{1}{\beta} \ln \left [
  1 + e^{-\beta \mu_0} 
  \right ],
\label{eq:et}
\end{equation} 
where $\mu_0(T)$ is the chemical potential and at $T=0$ $\mu_0 = \ve_F$.
With the 2D parabolic energy dispersion $\ve = (\hbar k_F)^2/2m$
Eq.~(\ref{eq:sigma0}) becomes the 2D
conductivity formula, $\sigma = ne^2 \langle \tau \rangle /m$.

With the classical average velocity $\langle v_x \rangle$ the current
density is given by
\begin{equation}
J_x = n(T)e\langle v_x \rangle,
\label{eq:j}
\end{equation}
where we use the total electron density at finite $T$  instead of the
zero temperature density $n_0$, and $n(T)$ is given by
\begin{equation}
n(T) = \int D(\ve) f(\ve) d\ve.
\label{eq:nt}
\end{equation}
From Eqs.~(\ref{eq:j0}) and (\ref{eq:j}) we have
\begin{equation}
\langle v_x \rangle = \frac{\sigma(T)}{e n(T)} E_x.
\end{equation}
Then the mobility can be defined by
(we note that we have used the standard notation $\mu$ to imply both
mobility and chemical potential which should not cause any confusion
since they do not arise in the same equation in the text and it should
be clear from the context whether mobility or chemical potential is
being discussed) 
\begin{equation}
\mu(T) = \frac{\langle v_x \rangle}{E_x} = \frac{\sigma(T)}{e n(T)}.
\label{eq:mut}
\end{equation}

Now we define the mean free path from the average scattering time as
\begin{equation}
l(T) = v_F \langle \tau \rangle = {\sigma}\frac{h}{e^2} \frac{2}{g}
\frac{\hbar v_F}{\ve(T)}. 
\label{eq:lt}
\end{equation}
In mobility [Eq.~(\ref{eq:mut})] and mean free path
[Eq.~(\ref{eq:lt})] the total density and 
thermal energy at finite temperature are used instead of $n_0$ and
$\ve_F$, i.e.,
\begin{equation}
\mu(T) = \frac{\sigma(T)}{e n_0}.
\label{eq:mu0}
\end{equation}
and
\begin{equation}
l(T) = {\sigma}\frac{h}{e^2} \frac{2}{g}
\frac{\hbar v_F}{\ve_F}. 
\label{eq:l0}
\end{equation}

\begin{figure}[ht]
	\centering
	\includegraphics[width=1\columnwidth]{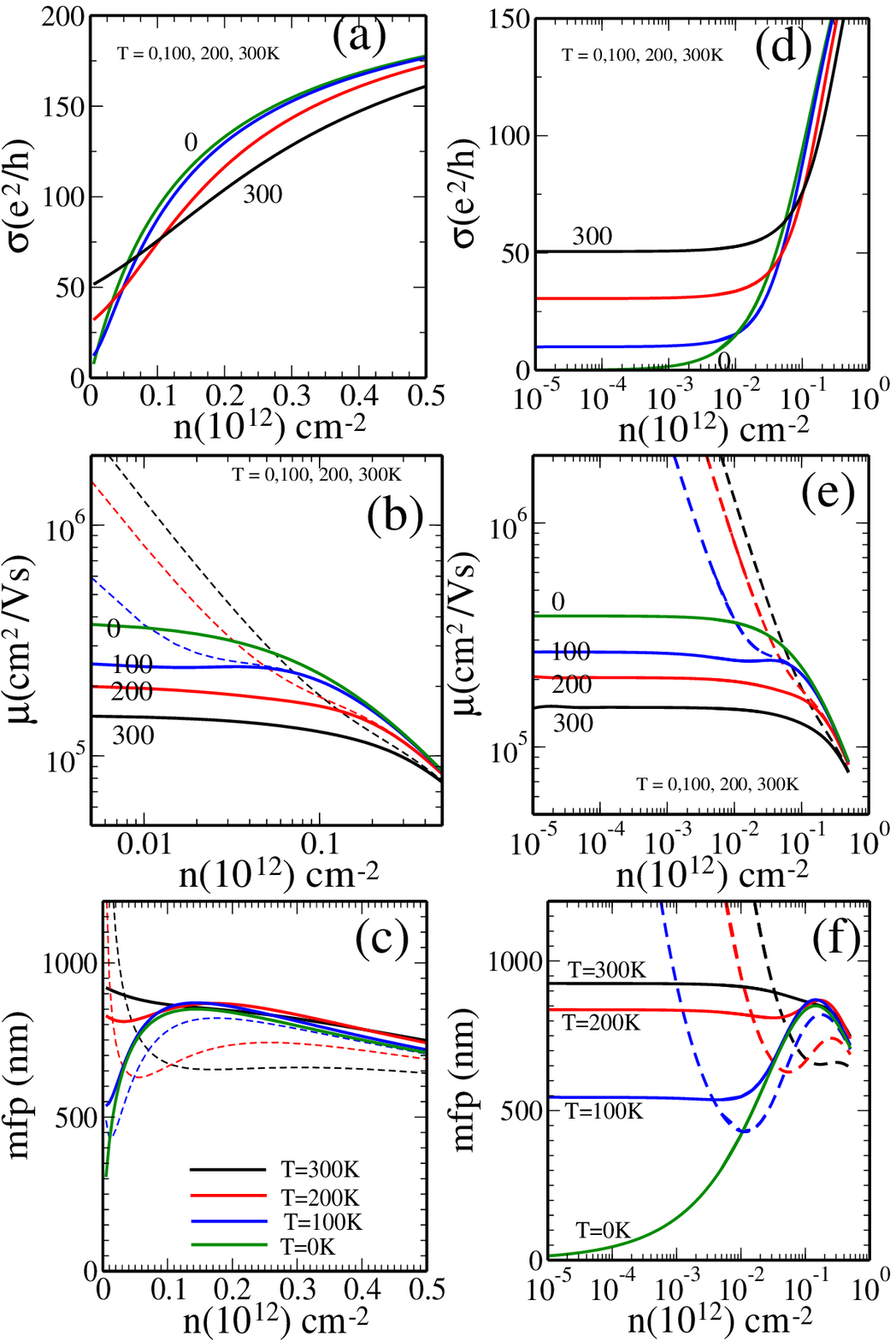}
	\caption{
(color online) Conductivity of suspended graphene corresponding to the experimental
data of Bolotin {\it et al.}\cite{bolotin2008}. 
(a) The calculated conductivity as a function
          of density for different 
          temperatures $T=0,$ 100, 
200, 300K (top to bottom) with $n_i=0.85\times 10^{10}$  
cm$^2$ and $n_{\delta}V_{\delta}^2 = 1.5$ (eV\AA)$^2$. 
In (b) and (c) the mobility and mean free path are shown, respectively.
The solid lines are calculated with temperature dependent $n(T)$ and
$\ve(T)$, and the dashed lines are calculated with a zero temperature density
$n_0$ and the energy $E_F$.
In (d), (e), and (f) we show $\sigma$, $\mu$, and $l$ at low densities
(down to $n = 10^7 cm^{-2}$), respectively. Note that as $n\rightarrow 0$ or $T/T_F
\rightarrow \infty$ for a fixed temperature, $\mu(T) \propto
\sigma(T)/T^2$ and $l(T) \propto \sigma(T)/T$. Thus, both $\mu(n\rightarrow
0)$ and $l(n\rightarrow 0)$ saturate at a finite temperature.
}
\label{fig:fig2}
\end{figure}

We note as mentioned already that there are two possible alternative
definition above for the mobility $\mu$ and mean free path $l$,
depending on whether one uses the gate induced doping density $n_0$ or
the full carrier density $n(T)$ including the thermally excited
carriers. For $T<T_F$ where $T_F$ ($=\epsilon_F/k_B$) is always
defined with respect to the $T=0$ carrier density induced by the
gate, the two definitions are equivalent since $n(T) \approx n_0 =
n$. But, at very high temperatures (or very low doping density), $n(T)
\gg n_0$ because $T \gg T_F$. For ordinary graphene on substrates,
where the puddle-induced density inhomogeneity introduces a cut-off
density of $n_c\sim 10^{12}$ cm$^{-2}$ with a corresponding $T_F \sim
1250 K$, by definition $n(T) = n_0 = n$, and there is not much of a
difference between the two different ways of defining the mobility and
the mean free path. But, for suspended graphene, where $n_c$ is small
and thus very small values of $n_0$ are meaningful,
the two distinct ways of defining $\mu$ and $l$ make a big difference
for small external doping density ($n_0 \sim 0$), i.e., near Dirac
point.

To make the above point more explicit, we show in Fig.~\ref{fig:fig1}
our  calculated $n=n(T)$ and $\ve(T)$ compared with $n_0$ and
$\ve_F$, respectively. It is clear that for $T/T_F \agt 1$, there
is a substantial difference between $n(T)$ and $n_0 = n(T=0)$. Since
$T_F \sim 12 K$ for $n_0 \sim 10^8$ cm$^{-2}$, close to the Dirac
point, the two quantities ($n$ and $n_0$) may differ a lot even at
moderate temperatures.

\begin{figure}[ht]
	\centering
	\includegraphics[width=1\columnwidth]{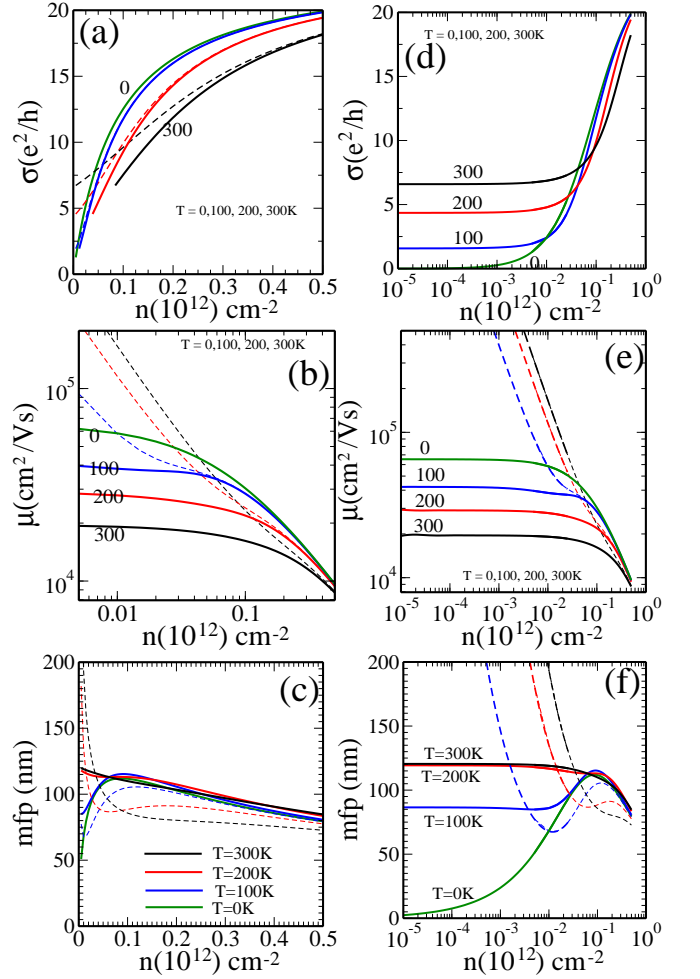}
	\caption{
(color online) Conductivity corresponding to the experimental
data of Du {\it et al.}\cite{du2009}. 
(a) The calculated conductivity as a function of density for different 
temperatures $T=0,$ 100, 200, 300K (top to bottom) with $n_i=5.0\times 10^{10}$ 
cm$^2$ and  $n_{\delta}V_{\delta}^2 = 14.7$ (eV\AA)$^2$. 
In (b) and (c) mobility and mean free path are shown as a function
of density, respectively.
The solid lines are calculated with temperature dependent $n(T)$ and
$\ve(T)$, and the dashed lines are calculated with a zero temperature density
$n_0$ and an energy $E_F$.
In (d), (e), and (f) we show $\sigma$, $\mu$, and $l$ at low
densities respectively. 
}
\label{fig:fig3}
\end{figure}

\begin{figure}[ht]
	\centering
	\includegraphics[width=1\columnwidth]{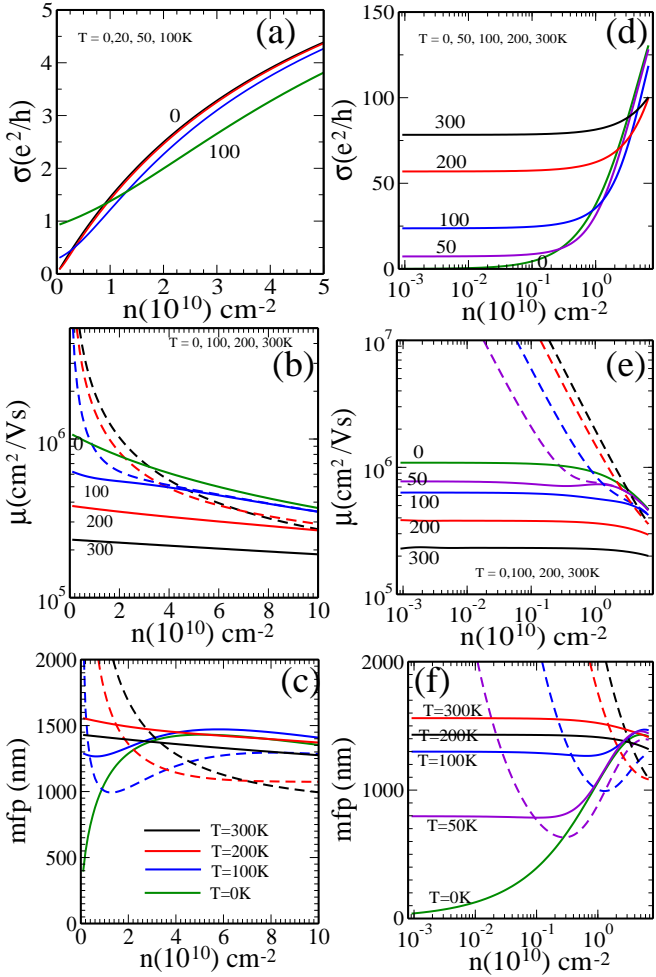}
	\caption{
(color online) Conductivity corresponding to the experimental
data of Mayorov {\it et al.}\cite{mayorov2012}.
(a) The calculated conductivity as a function
          of density for different 
          temperatures $T=0,$ 100, 
200, 300K (top to bottom) with
$n_i=0.3\times 10^{10}$ 
cm$^2$ and  $n_{\delta}V_{\delta}^2 = 1.5$ (eV\AA)$^2$. 
In (b) and (c) mobility and mean free path are shown as a function
of density, respectively.
The solid lines are calculated with temperature dependent $n(T)$ and
$\ve(T)$, and the dashed lines are calculated with a zero temperature density
$n_0$ and an energy $E_F$.
In (d), (e), and (f) we show $\sigma$, $\mu$, and $l$ at low densities,
respectively. 
}
\label{fig:fig4}
\end{figure}

\begin{figure}[t]
	\centering
	\includegraphics[width=1.0\columnwidth]{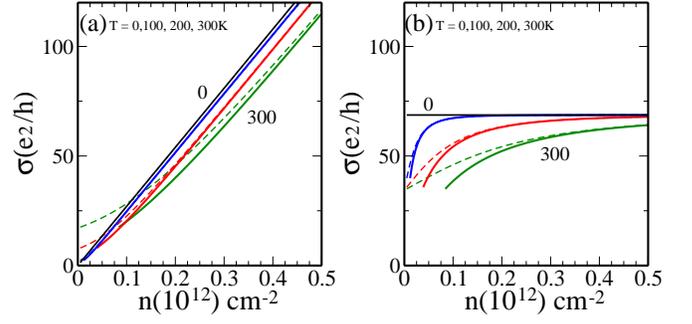}
	\caption{
(color online) Calculated conductivity as a function of density for different
temperatures: (a) for long-range Coulomb potential with
$n_i=10^{10}cm^{-2}$ and (b) for neutral short range potential with
$n_{\delta}V_{\delta}^2 = 5 (eV\AA)^2$.
Solid lines indicate $\sigma$ vs. $n(T)$, and dashed lines indicate
$\sigma$ vs. $n_0=n(T=0)$ or density induced by only gate voltage. 
}
\label{fig:fig5}
\end{figure}

\begin{figure}[t]
	\centering
	\includegraphics[width=1.0\columnwidth]{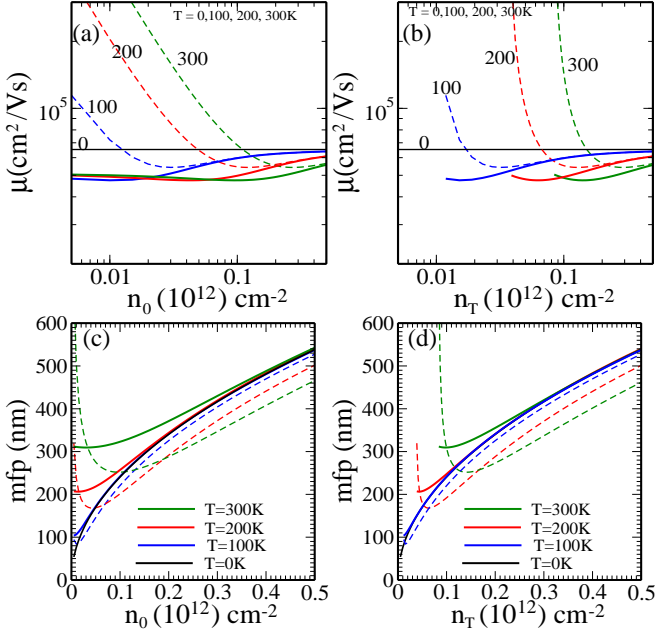}
	\caption{
(color online) (a) and (b) Calculated mobility as a function of
          density for different  
temperatures and for long-range Coulomb potential with
$n_i=10^{10}cm^{-2}$. Here $n_0$ indicates the density induced by the
gate voltage and $n(T)$ indicates the total density, i.e. the density
from gate plus the density from thermal excitations. Solid lines
represent Eq.~(\ref{eq:mut}) with $n(T)$ and 
dashed lines represent Eq.~(\ref{eq:mu0}) with $n_0=n(T=0)$ or density induced
by only gate voltage.  
In (c) and (d) the mean free path are shown as a function of density.
Solid (dashed) lines indicate Eq.~(\ref{eq:lt}) [Eq.~(\ref{eq:l0})].
}
\label{fig:fig6}
\end{figure}

\begin{figure}[t]
	\centering
	\includegraphics[width=1.0\columnwidth]{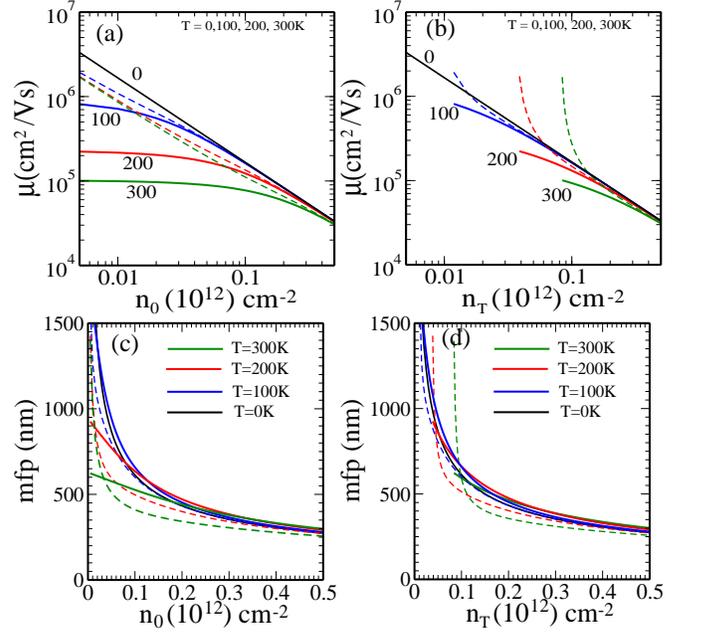}
	\caption{
(color online) (a) and (b) Calculated mobility as a function of
          density for different  
temperatures and for short-range neutral potential with
$n_{\delta}V_{\delta}^2 = 5 (eV\AA)^2$. Solid lines represent Eq.~(\ref{eq:mut}) with $n(T)$ and
dashed lines represent Eq.~(\ref{eq:mu0}) with $n_0=n(T=0)$ or density induced
by only gate voltage.  
In (c) and (d) the mean free path are shown as a function of density.
Solid (dashed) lines indicate Eq.~(\ref{eq:lt}) [Eq.~(\ref{eq:l0})].
}
\label{fig:fig7}
\end{figure}

\begin{figure}[ht]
	\centering
	\includegraphics[width=1\columnwidth]{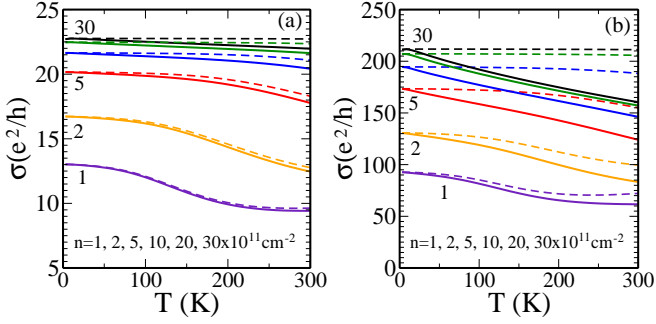}
	\caption{
(color online) Temperature dependent conductivity of suspended graphene corresponding
to the experimental data of (a) Du {\it et al}.\cite{du2008} and (b) Bolotin
{\it et al.}\cite{bolotin2008}. The same parameters used in
Figs.~\ref{fig:fig2} and \ref{fig:fig3} are used in 
this calculation.  The solid (dashed) lines indicate the results with
(without) phonon scattering.  
}
\label{fig:fig8}
\end{figure}

\begin{figure}[ht]
	\centering
	\includegraphics[width=1\columnwidth]{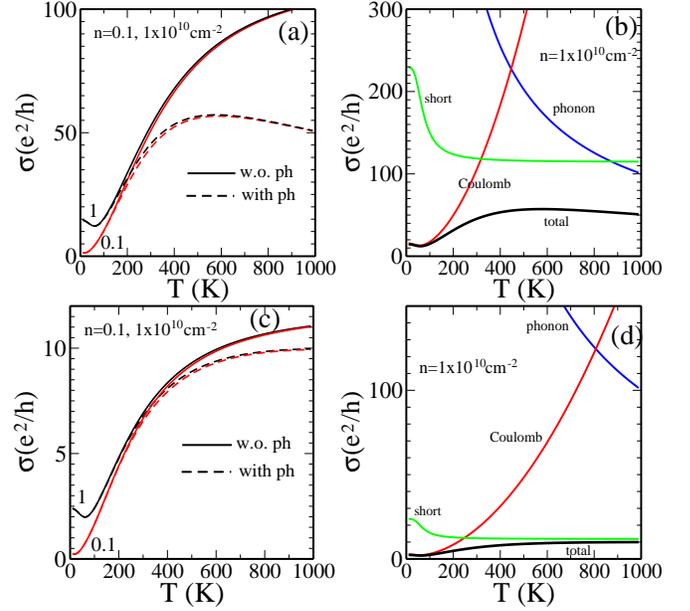}
	\caption{
(color online) Temperature dependent conductivity of suspended graphene corresponding
to the experimental data of (a)(b) Bolotin {\it et al}.\cite{bolotin2008} and
(c)(d) Du 
{\it et al.}\cite{du2008}. The same parameters used in
Figs.~\ref{fig:fig2} and \ref{fig:fig3} are used in 
this calculation.  In (a)(c) the solid (dashed) lines indicate the
results without (with) phonon scattering, and black (red) lines
indicate the results for a density $n=10^{10} cm^{-2}$ ($n=10^{9}cm^{-2}$).  
In (b) (d) the individual contribution to the total conductivity (black
line) is shown for a density $n=10^{10}cm^{-2}$. In (d)
the non-monotonic behavior at high density does not appear due to the
strong short-range potential scattering, but in high mobility 
samples (b) the non-monotonic behavior shows up due to the much weaker
neutral impurity scatterings. 
}
\label{fig:fig9}
\end{figure}

In Figs.~\ref{fig:fig2}--\ref{fig:fig9}, we show our calculated SG
transport properties as functions of density and temperature
neglecting all effects of density inhomogeneity or puddles -- our
theory should therefore be cut off at some very low doping density
($\alt 10^9$ cm$^{-2}$) where puddles become relevant in high-quality
SG. In Figs.~\ref{fig:fig2}---\ref{fig:fig7}, we show density
dependence for a few representative temperatures whereas in
Figs.~\ref{fig:fig8} and \ref{fig:fig9} we show the calculated
temperature dependence for a few fixed doping densities. We include
phonon effects only in Fig.~\ref{fig:fig8} and \ref{fig:fig9} since
our main interest is low-temperature transport. In obtaining our
numerical results, we focus on three published experimental SG works
in the literature: Bolotin {\it et al.}
\cite{bolotin2008,bolotin2009}, Du {\it et al.} \cite{du2008,du2009},
and Mayorov {\it et al.}\cite{mayorov2012}. Our
goal is not data fitting or getting precise agreement with the
experimental results since the details of disorder are unknown in the
experimental systems, and we cannot rule out the possibility of
additional scattering mechanisms not included in our model being
operational in the experimental systems. What we are interested in is
obtaining the broad qualitative features of SG transport data in our
theory in order to critically assess the issues of ballistic versus
diffusive SG transport and the access to the Dirac point at low
carrier density. We mention that the authors in all three of these
works \cite{du2008,du2009,bolotin2008,bolotin2009,mayorov2012}
interpret their data in terms of ballistic transport mainly by 
comparing their extracted mean free paths with the sample size. We
will critically examine the nature (ballistic or diffusive) of
transport in these experiments.

For each experiment, we choose a set of disorder parameters as shown
below based on the best overall semi-quantitative and qualitative
agreement with the data, keeping these disorder parameters fixed for
all the presented results. The acoustic phonon scattering parameters
are standard and are taken to be $D=19$ eV; $\rho_m =7.6\times
10^{-8}$ g/cm$^2$; $v_{ph}=2\times10^6$ cm/s. We
use the following disorder parameters for each experiment.
\[
{\rm Du} \; et \; al. \;\; n_i = 6.5\times 10^{10} cm^{-2}, \; n_{\delta} V_{\delta}^2 =
15.7 (eV\AA)^2;
\]
\[
{\rm Bolotin} \; et \;al. \;\; n_i = 1.2 \times 10^{10} cm^{-2}, \;
n_{\delta}V_{\delta}^2 = 1.5 (eV\AA)^2;
\]
\[
{\rm Mayorov} \; et \; al. \;\; n_i = 0.3 \times 10^{10} cm^{-2}; \; n_{\delta}
V_{\delta}^2 = 1.5(eV\AA)^2. 
\]
We note that consistent with the experimental SG sample quality, our
disorder is the strongest (weakest) in Du (Mayorov) with Bolotin
disorder being intermediate. This is consistent with the claimed
high-density mobility being $\sim 5000,000$ cm$^2$/Vs, $\sim 200,000$
cm$^2$/Vs, and $\sim 100,000$ cm$^2$/Vs, respectively, in the three
experiments (although the precise value of the sample mobility may not
be a meaningful quality since the mobility depends on both density and
temperature).

In Fig.~\ref{fig:fig2}---\ref{fig:fig4}, we show our calculated
$\sigma(n)$, $\mu(n)$, and $l(n)$ as a function of doping density $n$
(alluded to $n_0$ above) for $T=0$, 100, 200, 300K for the three
experimental SG samples, respectively -- we emphasize that for small
values of $n_0$, where $T/T_F>1$ condition may apply, the alternative
definitions for the chemical potential and the mean free path would
lead to large quantitative differences since, as is obvious from
Fig.~\ref{fig:fig1}, $n(T) \gg n_0$ in this regime.

Our calculated $\sigma(n)$ results for the three experimental samples
in Figs.~\ref{fig:fig2}--\ref{fig:fig4} manifest similar qualitative
behavior with large quantitative differences because of the differences
in the details of the underlying disorder. In particular, the
following salient features of the results are consistent with the
experimental findings in high-quality SG samples: (1) $\sigma(n)$
manifests sublinear density dependence, simulating $\sigma \sim
\sqrt{n}$, over an extended density range, thus calling into question
the experimental interpretation of SG transport being ballistic
\cite{muller2009} based
entirely on this sublinear density dependence; (2) at the lowest density,
$\sigma(n)$ is always limited by the long-range Coulomb scattering
(with $\rho \propto n$), but the competition between long-range and
short-range disorder (which leads to the effective sublinear density
dependence over an extended density range) and the existence of the
low-density puddle-dependent cut-off (not included in the current
theory) may mask this linear density dependence in high-quality SG
samples where random charged impurity disorder is presumably rather
low; (3) at the lowest density, the system always would manifest
insulating temperature dependence because of the dominance of the
thermal excitation in the gapless system [this is obvious in the panel
(d) of Figs.~\ref{fig:fig2}--\ref{fig:fig4}] near the Dirac point --
there is a density-dependent crossover to the metallic behavior at
higher carrier densities, emphasizing that the characteristic Dirac
point insulating transport behavior is a high-temperature crossover
behavior (which may not be apparent for $T/T_F \ll 1$; (4) the
calculated mobility approaches $\sim 5 \times 10^5$, $10^5$, and
$10^6$ cm$^2$/Vs respectively in the Bolotin, Du, and Mayorov samples
close to the Dirac point, showing the unprecedentedly high qualities of
these SG systems; (5) it is misleading to characterize the mobility
(or the mean free path) using the gate induced density since this
would produce erroneously large mobility and mean free path at low
gate voltages, and in fact, would imply a divergent mobility (or mean
free path) at the Dirac point -- when the full density $n(T)$ is used
in defining the mobility (or mean free path), the low-density mobility
and mean free path saturate providing the correct characterization;
(6) both definitions [using $n_0$ or $n(T)$] give
identical mobility and mean free path values for high carrier
densities ($\agt 10^{12}$ cm$^{-2}$), as one expects since $n(T) \sim
n_0$ for high densities; (7) although broadly in qualitative agreement
with the experimental data, there are important discrepancies between
our theory and experiment in the details, most likely
because of our neglect of other possible scattering mechanisms in the
experimental systems; (8) the appropriate mean free path (at low
densities near the Dirac point) varies between $\sim 100$nm (Du
sample) and $\sim 1000$nm (Mayorov sample), and therefore true
ballistic transport measurements would require sample size $< 0.1
\mu$m, and one must observe the sample length dependent conductivity
to validate any ballistic transport behavior.

Since long-range and short-range disorders affect $\sigma(n,T)$
qualitatively differently (and the nature of disorder in the
experimental samples is not known based on any independent
measurements), we depict in Figs.~\ref{fig:fig5}--\ref{fig:fig7} the
distinct theoretical dependence of conductivity on the long-range and
short-range disorder separately (in contrast to
Figs.~\ref{fig:fig2}--\ref{fig:fig4} where both are included together in
the theory) on the density for the SG system. To bring out the
qualitatively different dependence of conductivity, mobility, and mean
free path on $n_0$ (the doping density) or $n(T)$ at low density, we
show in Figs.~\ref{fig:fig5}--\ref{fig:fig7} dependence on both $n_0$
and $n(T)$ separately. In Fig.~\ref{fig:fig5}, we show the calculated
conductivity  whereas in Figs.~\ref{fig:fig6} and \ref{fig:fig7} we
show the mobility and the mean free path. In Fig.~\ref{fig:fig5} we
show $\sigma(n)$ for just long-range disorder or just short-range
disorder with the two different density dependences [$n_0$ or $n(T)$]
showing quantitative differences only at low values of $n$ (or
equivalently, high values of $T$) with the two being identical (by
definition) at $T=0$ since $n(T=0) \equiv n_0$. In Figs.~\ref{fig:fig6}
and \ref{fig:fig7}, we show the calculated mobility and mean free path
for long-range (Fig.~\ref{fig:fig6}) and short-range
(Fig.~\ref{fig:fig7}) disorder with each case also providing the
dependence on $n_0$ and $n(T)$. The important qualitative conclusion
from Figs.~\ref{fig:fig5}--\ref{fig:fig7} is that one should always
extract mobility and mean free path using the correct total density
$n(T)$ rather than just the doping density $n_0$, particularly at low
carrier densities because the extracted mobility and mean free path
for the two definitions differ qualitatively as the Dirac point is
approached with the distinction between the two definitions being
much larger for long-range disorder. Our work establishes that derived
quantities such as mean free path and mobility, which involve an
effective division of the experimentally measured conductivity by a
density, are not meaningful for graphene (particularly at low
densities, approaching the Dirac point) because $n(T)/n_0$ diverges at
the Dirac point. This is not a serious problem for graphene on
substrates because the puddle-induced cut-off density $n_c$ ensures
that $n(T) \approx n_0 \approx n$, but in high quality suspended
graphene mobility and mean free path are meaningful only if they are
extracted at high density where $n(T) \approx n_0$.

All the above results (Figs.~\ref{fig:fig2}--\ref{fig:fig3}) ignore
phonon effects which are very weak in graphene and only affect high
temperature transport. In Figs.~\ref{fig:fig8} and \ref{fig:fig9}, we
show the explicit effects of acoustic phonon scattering in the theory
by comparing results for $\sigma(T)$ including and excluding phonons
in the calculation at high (Fig.~\ref{fig:fig8}) and low
(Fig.~\ref{fig:fig9}) carrier density, and for the Bolotin {\it et
  al.}\cite{bolotin2008,bolotin2009} and Du {\it et al.}\cite{du2008,du2009}
samples. In general, the phonon scattering effect is much stronger for
the Bolotin {\it et al.} sample than the Du {\it et al.} sample
because of the much higher 
quality (lower disorder) of the former. 
This finding is completely consistent with our theoretical analysis in
Section II and III where we establish that the Dirac point
conductivity would be affected by phonons even at rather low
temperatures for very clean samples with low values of $n_i$.
Our basic finding is that
phonons introduce metallic temperature dependence at higher carrier
density nullifying the intrinsic insulating temperature dependence
arising from Coulomb disorder, but in general the insulating
temperature dependence remains quite strong upto the room temperature
at low carrier density (Fig.~\ref{fig:fig9}) in high-quality suspended
graphene. We note that both Figs.~\ref{fig:fig8} and ~\ref{fig:fig9}
clearly show the very low-temperature ($T/T_F \ll 1$) weak metallic
T-dependence of $\sigma(T)$ arising entirely from the Fermi surface
effect which is more strongly manifested in the higher density
(Fig.~\ref{fig:fig8}) system. This again reinforces our claim that the
insulating behavior of $\sigma(n,T)$, which is the hallmark of the
Dirac point transport property, is much better studied as a high
temperature phenomenon in low-density SG. This insulating behavior has
clearly been observed by Bolotin {\it et al.}, Du {\it et al.}, and
Mayorov {\it et al.}, establishing that
all three SG samples are reflecting intrinsic Dirac point transport
behavior in their very high quality SG samples. Based on our results
we contend that the observation of low density power-law insulating
temperature dependence in graphene is a direct manifestation of the
Dirac point behavior.

\section{conclusion}

We have provided in this work a detailed theoretical study of the
density and temperature dependent conductivity of low-disorder
suspended graphene within the semiclassical Drude-Boltzmann transport
theory neglecting density inhomogeneity (i.e. puddle) effects. Our
theory includes three independent scattering mechanisms: long-range
Coulomb disorder, short-range $\delta$-function disorder, and acoustic
phonon scattering. We establish, by comparing our detailed numerical
results for the conductivity with three recent experimental
studies\cite{du2008,du2009,bolotin2008,bolotin2009,mayorov2012}, 
that the measured low-density conductivity in existing experiments on
suspended graphene is approaching at least some aspects of the
intrinsic Dirac point behavior.

Some of our more important qualitative conclusions are: (1) the
intrinsic Dirac point behavior is better manifested at higher (lower)
temperatures (densities) staying above the puddle-induced
characteristic density; (2) the observation of a power law insulating
temperature dependence of conductivity is a direct manifestation of
the Dirac point behavior; (3) at low doping densities, it is not
meaningful to characterize the system using derived quantities
(e.g. mobility or mean free path) because of the considerable
ambiguity in which density (just the extrinsic doping density or the
total density including thermal excitations) should be used in the
definition of mobility (or mean free path); (4) the competition among
long-range and short-range disorder plus phonon scattering could lead
to complex (and even non-monotonic) dependence of the conductivity on
temperature and density, and it is not meaningful to conclude about
the underlying nature of the transport behavior (ballistic or
diffusive; localized or extended, etc.) based just on preconceived
notions about the expected density and temperature dependence for
various processes; (5) by improving sample quality and reducing
disorder, it should be possible to approach the Dirac point
indefinitely through careful conductivity measurements in suspended
graphene, providing unique opportunities to study in the future many
interesting effects not included in our theory (e.g, interaction,
localization, ripple, flexural phonons); (6) phonons could affect the
Dirac point conductivity in high-quality SG down to arbitrarily low
temperatures since the Bloch-Gr\"{u}neisen temperature becomes vanishingly
small near the Dirac point -- whether phonon effects will overcome the
insulating temperature dependence due to Coulomb disorder depend on
the details of the amount of disorder scattering effective in the
system.

We conclude by emphasizing that our results establish that how close
in density one has approached the Dirac point can be estimated by
seeing how high in temperature the Coulomb disorder induced insulating temperature dependence
persists in a particular graphene sample
(or paradoxically, how low in temperature the acoustic phonon effects
persist if the graphene sample is devoid of Coulomb disorder causing
the insulating behavior).

\section*{acknowledgments}
This work is supported by US-ONR.


\end{document}